\documentclass[journal, final, 10pt, twocolumn]{IEEEtran}
\usepackage{layout}
\usepackage[letterpaper, left=.92in, right=.92in, bottom=.7in, top=0.70in]{geometry}

\usepackage{graphicx}
\usepackage{amsmath}
\usepackage{subfig}
\usepackage{mathtools}
\usepackage{epsfig}
\usepackage{float}
\usepackage{lipsum}
\usepackage{stfloats}
\usepackage{array}
\usepackage{amssymb}
\usepackage{cite}
\usepackage{url}
\usepackage{hyperref}
\usepackage{amsthm}
\usepackage{algorithm}
\usepackage{algorithmic}
\usepackage{multirow}
\usepackage{upgreek}
\usepackage{diagbox}
\usepackage{dsfont}
\usepackage{xifthen}% provides \isempty test
\usepackage{xcolor}
\usepackage[colorinlistoftodos,textsize=tiny]{todonotes}

\normalsize
\allowdisplaybreaks

\newcommand{\GD}[1][]{
\ifthenelse{\isempty{#1}}%
{\text{G}_{i,j}^k}% if #1 is empty
{{(\text{G}_{i,j}^k)}^{#1}}% if #1 is not empty
}

\theoremstyle{definition}

\usepackage{layout}
\usepackage{bm}
\usepackage{comment}
\usepackage{color,soul}
\usepackage{makecell}
\usepackage{amssymb}

\usepackage{subfig}
\usepackage{tikz}
\usetikzlibrary{positioning}
\usetikzlibrary{calc,through,backgrounds}
\def\*#1{\mathbf{#1}}
\def\w*#1{\widehat{#1}}

\def\s*#1{\mathsf{#1}}
\def\S*#1{\bm{\mathsf{#1}}}

\def\c*#1{\mathcal{#1}}
\def\C*#1{\bm{\mathcal{#1}}}
\def\T*#1{\text{#1}}
\def\*#1{\mathbf{#1}}
\begin{document}

\title{Video on Demand Streaming Using RL-based Edge Caching in 5G Networks}
\author{\IEEEauthorblockN{Rasoul Nikbakht},  \IEEEauthorblockN{Sarang Kahvazadeh},  \IEEEauthorblockN{Josep Mangues-Bafalluy} \\
	\IEEEauthorblockA{Centre Tecnològic Telecomunicacions Catalunya (CTTC)\\
		08860 Castelldefels, Spain.\\
		Email: \{rnikbakht, skahvazadeh, jmangues\}@cttc.es}
}

\maketitle

\begin{abstract}
Edge caching can significantly improve the 5G networks' performance both in terms of delay and backhaul traffic. We use a reinforcement learning-based (RL-based) caching technique that can adapt to time-location-dependent popularity patterns for on-demand video contents. In a private 5G, we implement the proposed caching scheme as two virtual network functions (VNFs), edge and remote servers,  and measure the cache hit ratio as a KPI. Combined with the HLS protocol, the proposed video-on-demand (VoD) streaming is a reliable and scalable service that can adapt to content popularity.  

\end{abstract}
%------------------------------------------------------------
%------------------------------------------------------------
\section{Introduction}
It is expected that by 2022, media content will account for more than $80\%$ of Internet traffic, and popular media will make up the majority of the media traffic \cite{cisco}. This naturally opens the door for caching and storage in the cloud, as media files are usually large and frequently accessed. 
Content Delivery Network (CDN) services fill this gap by delivering content to users over the Internet. CDNs cache and deliver the requested content from a remote server, allowing users to load the content faster. However, for new emerging applications such as on-site media production and low latency live streaming, the cache server should be located even closer to the end-user, so the need for edge-caching becomes apparent \cite{5g_appliation_2019,shuja2020applying,CaaS_2017}.
 
We implemented edge-caching concept as virtual cache servers (VCaches) in the 5G infrastructure. Each Vcache is a virtual network function (VNF) that can be controlled by an open source management and orchestration (OSM) software stack. 

%The proposed approach capitalizes on:
%\begin{itemize}
%    \item Scalable and effcient implementation and testing in a private 5G network using OSM framework.% The implementation uses Nginx as an entry point, Flask as an application server, and HLS protocol for video playback over Internet. The HLS playlist is pre-processed using FFmpeg, which is an open source tool for video and stream editing. 
%	\item An RL-based algorithm that can be trained using the content popularity data.  In this work, we use discrete version of the soft actor-critic networks \cite{haarnoja2018soft,christodoulou2019soft}.
%	%\item Simulated popularity pattern, where content popularity is modeled with a Zipf distribution. Nonetheless, the proposed framework can be extended to real-world data once it is in the production stage. 
%\end{itemize}

The main contributions of this work are as following:
\begin{itemize}
	\item An RL-based framework is used for edge caching. We take the number of requests for the cached content alongside their IDs as the state for the RL agent. Then, we use state-of-the-art soft actor-critic networks \cite{haarnoja2018soft,christodoulou2019soft} to obtain an efficient caching strategy. 
	\item We deploy an end-to-end 5G network with a core residing in the cloud and the cache server in the edge.
	\item Our solution is scalable, cloud-native, and can be automatically deployed on the network edge using the OSM framework.  
\end{itemize}

In \cite{DRL_based_caching2020_zhong}  centralized and decentralized caching using deep deterministic policy gradient is proposed. We adopt a similar workflow but with two major differences. First, we use the state-of-the-art soft actor-critic network for RL-based caching. Second, the state dimension in our proposed approach is smaller than the one in \cite{DRL_based_caching2020_zhong}, hence actor and critic networks are significantly simpler.
%Add a paragraph highlighting the advantages of the proposed approach (Production grade implementation, satisfactory caching results, open source code and components)

%The proposed RL-based caching is a scalable system that outperforms traditional caching techniques such as least frequency used (LFU), can adopt to varying content popularity, and is publicly available in \cite{github_repo}.
In this work, we use a commercial device (a smartphone or laptop) as a video playback tool and show the effectiveness of RL-based edge-caching for reducing delay and backhaul traffic. The proposed RL-based caching is a scalable system that outperforms traditional caching techniques such as least frequency used (LFU), can adapt to varying content popularity, and is built using fully open-sourced components. A containerized implementation of proposed video-on-demand streaming using RL-based caching is publicly available in \cite{github_repo}.

%In the following, we will go through the detail of RL-based caching and its implementation on a private 5G network.
%------------------------------------------------------------
%------------------------------------------------------------
\section{System architecture}
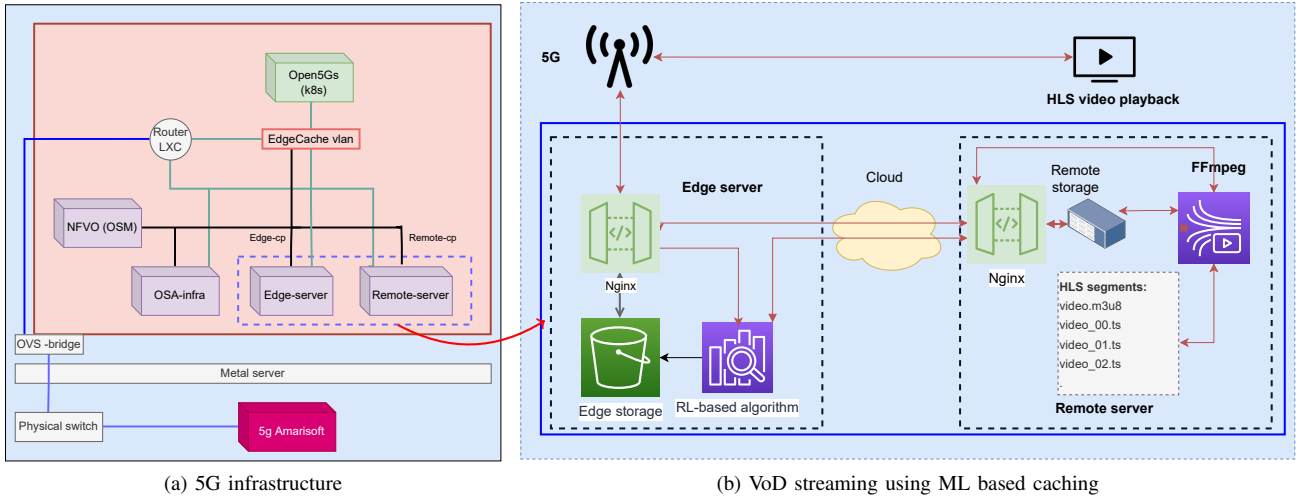
\begin{figure*}
	%	\centering
		\subfloat[][5G infrastructure]{
				\begin{tikzpicture}[remember picture]
				\node[inner sep=0pt] (5G_infra) at (0,0) {\includegraphics[width=.39\linewidth]{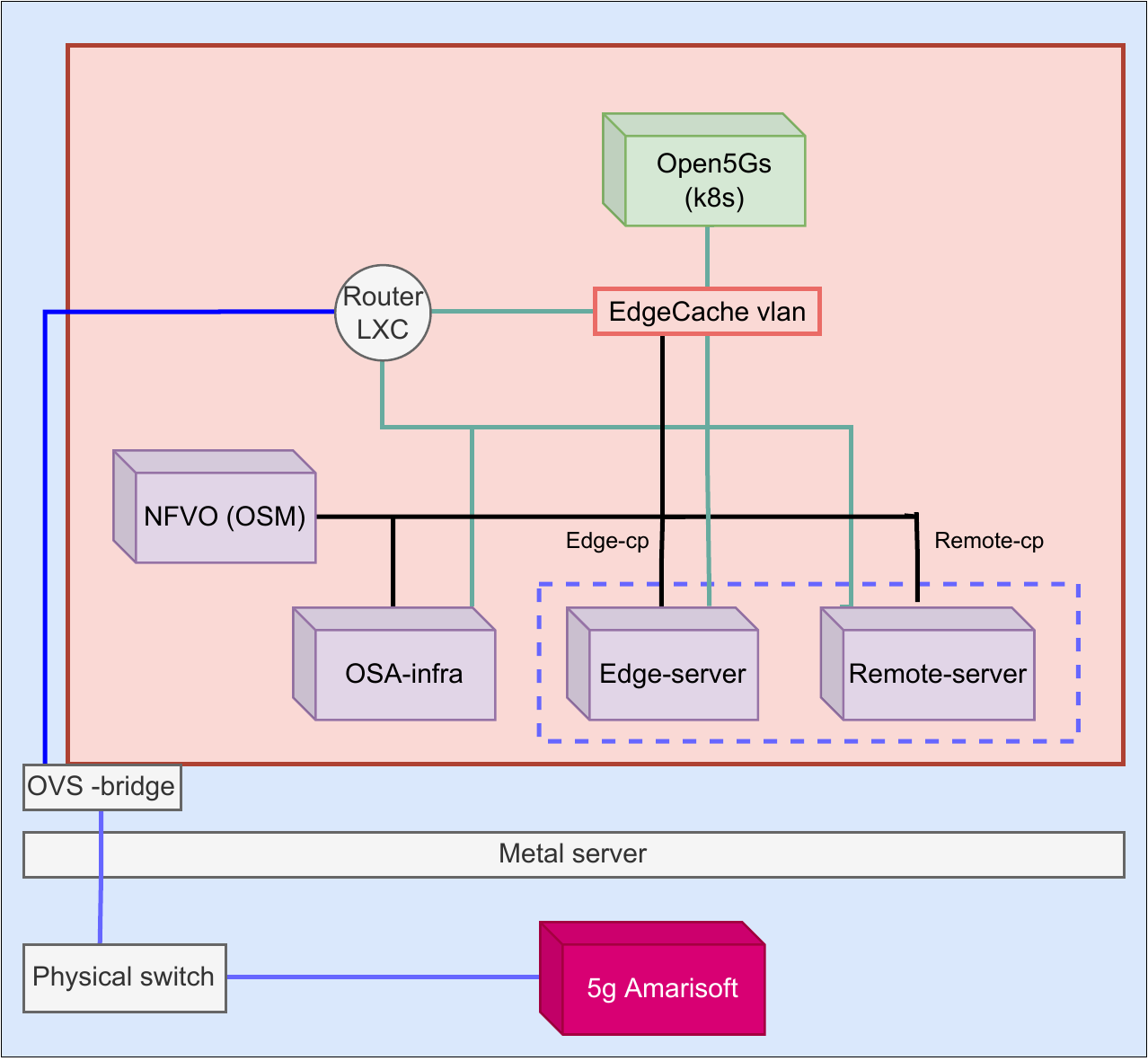}\label{fig:implementation_a}};
				 \node (A) at (1.8,-1.15) {}; % Point A
			    \end{tikzpicture}
				}
		\subfloat[][VoD streaming using ML based caching]{
				\begin{tikzpicture}[remember picture]
					\node[inner sep=0pt] (whitehead) at (8.5,0){\includegraphics[width=.61\linewidth]{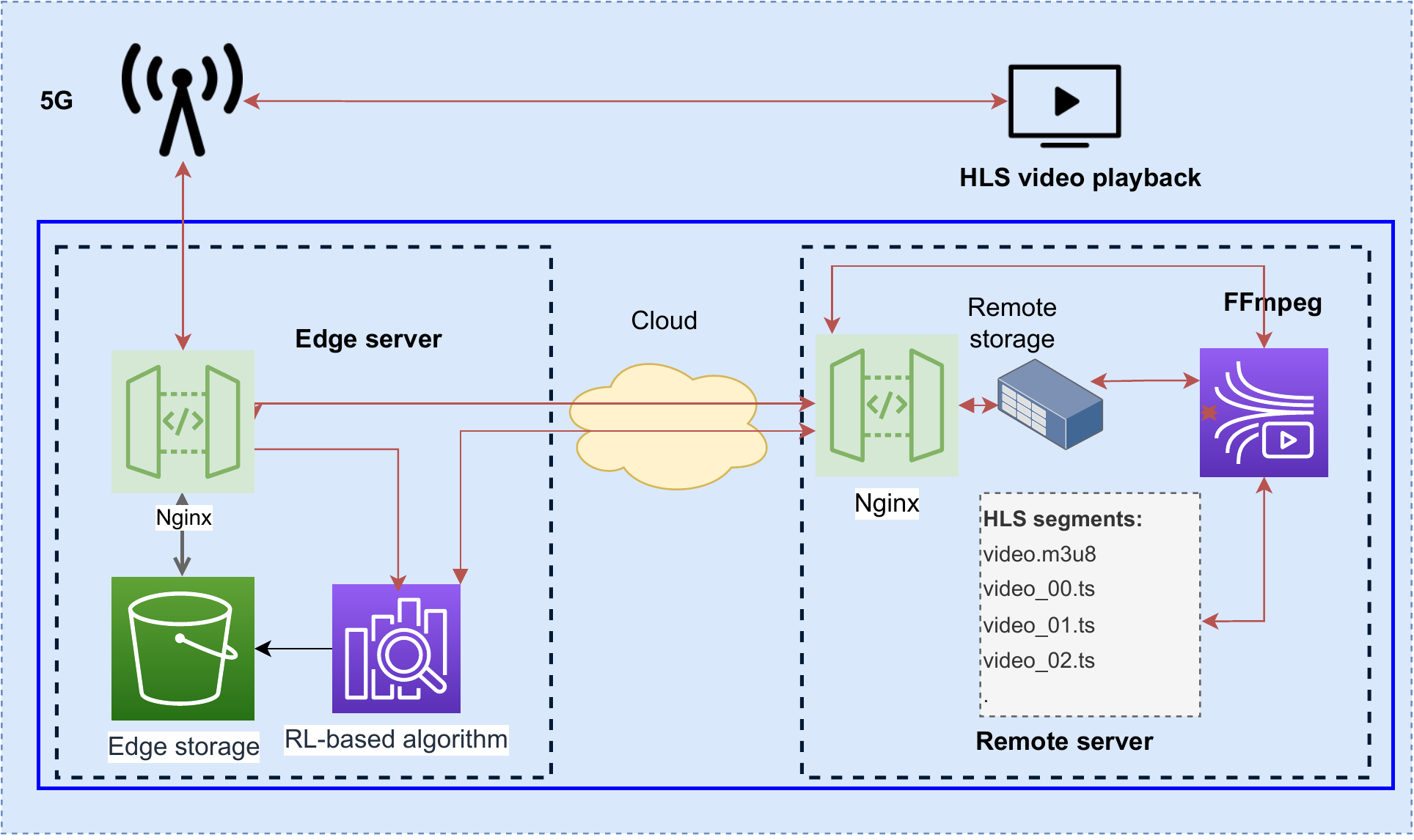}\label{fig:implementation_b}};
					\node (B) at (3.8,-1.1) {}; % Point B
				\end{tikzpicture}
		}
		\begin{tikzpicture}[overlay, remember picture]
		\path[->,red,thick] (A) edge [bend right] (B);
		%		\draw [-stealth](0,-5) -- (1,0);
	\end{tikzpicture}
	\caption{Implementation schematic}
	\label{fig:implementation}  

\end{figure*}

In this demonstration, we show how the RL-based edge-caching can be deployed, trained, and tested in a 5G network. Shown in Fig. \ref{fig:implementation} is a private 5G network alongside the proposed VoD streaming using RL-based edge caching. We use Open5GS \cite{open5gs} on the top of a Kubernetes cluster as core network function; OSM as orchestrator; Open-Stack-Ansible (OSA) and Linux container (LXC) as virtual infrastructure manager (VIM); and Amarisoft \cite{amarisoft} as radio access network and simbox for emulating user equipment.

The edge and remote servers are VNFs and launched using OSM alongside a network slice (NS). The OSM deployment is based on VNF and NS descriptors where the day zero operation for VNFs is conducted using the cloud.init script, which installs the required software, downloads the source code, and finally launches the remote and edge servers shown in Fig. \ref{fig:implementation_a}. In this implementation, we assumed an edge server located in the network edge and a remote server in the cloud. Since we don't have a dedicated server outside our organization (where we deployed 5G core), we implement both VNFs in the same infrastructure and introduce a reasonable random delay in the link between edge and remote servers. As a result, the user experiences an extra delay when being served by the remote server. In  Fig. \ref{fig:implementation_b}, we show this concept as the cloud. 

Illustrated in  Fig. \ref{fig:implementation_b} is the edge-cache implementation in the private 5G infrastructure, which consists of the following parts:
\begin{itemize}
	\item Edge server: The ML algorithm alongside a Nginx and edge storage have located in the edge server. These components are organized using Docker-compose and co-located in the same machine. 
	\item Remote server: It includes the FFmpeg, remote storage and an Nginx instance. The FFmpeg is controlled by Nginx and produces the HLS playlist. All the components again are implemented using Docker-compose.
\end{itemize}

The initial training for the ML algorithm is done during the Day 0 operation, but we can fine-tune or even retrain the ML algorithm from scratch upon the availability of the new observations (requests from users).

%---------------------------------------------------
%\subsection{Video-on-demand service}
%Th HLS protocol consists of a playlist and video segments. The user client asks for video playlist (video.m3u8) though 5G link. If content is cached (the playlist and video segments are available in the cache storage)
%\begin{itemize}
%	\item The Nginx in edge server receives the request and serves the user. In addition, the Nginx mirrors the user request to the ML algorithm.
%\end{itemize}
%If the content is not cached in the edge-server:
%\begin{itemize}
%	\item The edge server re-routes the user request to the remote server.
%	%\item The Nginx in the remote server first launches the FFmpeg script.	
%	%\item The FFmpeg script converts the video to an HLS playlist.
%	\item Nginx in the remote server directly serves the users. 
%	\item In the background, the application server that includes the RL-based caching algorithm updates the cached content on the edge-server (download the HLS playlist and video segments from the remote server)
%\end{itemize}
%-------------------------------------------------
\section{RL-based caching}
Let's assume we have a total of M contents of which C contents can be kept in a cache storage. Formulating the cache server as a RL algorithm, we define the state and action space of the RL, the logic of the edge caching, observation, and the reward function as the following.

\textit{State:}
\begin{itemize}
	\item Cached content IDs ($C_i$)
	\item The Total number of requests $R_i$ for each cached content in a sliding window of size $L$
\end{itemize}
The state S can be formulated as:
\begin{equation}
	S = \log \left[\left\{C_i\right\},\left\{R_i\right\}\right]\quad i \in \left\{\text{cached contents ID}\right\}
\end{equation}
Vector $S$ has a dimension of $2C$, and the $\log(.)$ operator makes sure that all the elements of the $S$, independent of their value, can impact the training of the RL agent. 

\textit{Action space:} The action space $A={0,1,...,C}$ has a dimension of $C+1$.
The DRL agent can either replace the selected cached content with the currently requested content for $A={1,2,..,C}$ or keep the cached contents the same for $A={0}$.
%\subsubsection{The logic of the edge caching}
%\begin{itemize}
%	\item If the content is cached, return the content (or its ID) and update the state.
%	\item If the content is not cached, then ask for the content from the server, and take an action.
%\end{itemize}

\textit{Observation:} It is a request for a file index. In this project, we assume that the observations follow the truncated Zipf distribution. It means that at any given time the content popularity has a specific structure that is imposed by Zipf distribution and its parameter. As an example, for the Zipf parameter of 1.25, a small percentage of the contents ($5\%$) accounts for the majority of the traffic ($80\%$). The proposed framework can be extended to real-world data once it is in the production stage.

\textit{Reward:} The cache hit probability is defined as
\begin{equation}
	P_h = \frac{\text{Cache hit in the sliding window $L$}}{L}
\end{equation}
For training the edge caching algorithm, the agent (edge cache) receives an observation and takes an action in a way that maximizes its long-term reward, which is the cache hit ratio.

\textit{Implementation:} We use soft actor-critic network for training the RL agent.  We build a custom Gym environment for modeling the RL-based caching agent and train it using Deep Reinforcement Learning Algorithms with PyTorch library \cite{github_repo_DRL}. 

%In the current format, the contents are full movies. In the follow up work, we are considering HLS segments as with different bit rate as a different content.  
Finally, the proposed RL-based caching approach:
\begin{itemize}
	\item Is flexible and can be applied to other types of popularity patterns or real-world data traffic.
	\item Can learn the time-location dependent popularity pattern.
	\item Can adapt to a sudden change in the popularity pattern (for example inside a music concert).
\end{itemize}

\section{Video-on-demand service}
Th HLS protocol consists of a playlist and video segments. The user client asks for video playlist (video.m3u8) though 5G link. If content is cached (the playlist and video segments are available in the cache storage), the Nginx in edge server receives the request and serves the user. In addition, the Nginx mirrors the user request to the ML algorithm.
%\begin{itemize}
%	\item The Nginx in edge server receives the request and serves the user. In addition, the Nginx mirrors the user request to the ML algorithm.
%\end{itemize}.

If the content is not cached in the edge-server, the edge server re-routes the user request to the remote server. Meanwhile, the application server that includes the RL-based caching algorithm updates the cached content on the edge-server (download the HLS playlist and video segments from the remote server) 
%\begin{figure*}
%	%	\centering
%%	\subfloat[][a]{\includegraphics[width=.48\linewidth]{Figs/5G_infrastructure_vod.drawio}\label{<figure1>}}
%%	\subfloat[][b]{\includegraphics[width=.48\linewidth]{Figs/Demo_paper.drawio.pdf}\label{<figure2>}}
%	\caption{Implementation schematic left: 5G infrastructure, right: VoD streaming using ML based caching}
%	\label{fig:implementation}  
%	\begin{tikzpicture}[remember picture]
%		\node[inner sep=0pt] (5G_infra) at (0,0) {\includegraphics[width=.47\linewidth]{Figs/5G_infrastructure_vod.drawio}\label{<figure1>}};
%		\node[inner sep=0pt] (whitehead) at (8.5,0){\includegraphics[width=.48\linewidth]{Figs/Demo_paper.drawio.pdf}\label{<figure2>}};
%		\path[->,red,thick] (1.8,-1.2) edge [bend right] (4.6,-1);
%%		\draw [-stealth](0,-5) -- (1,0);
%	\end{tikzpicture}
%\end{figure*}

%------------------------------------------------------------
%------------------------------------------------------------
\section{Demonstration}
Let's assume the introduced 5G network is up and running. Using OSM, we launch two VNFs for remote and edge servers and an NS for connecting them. Using a cloud-init file, OSM instantiates the VoD streaming service, which includes
\begin{itemize}
    \item Docker-composed-based application in remote and edge servers
    \item Initial training of RL-based caching
\end{itemize}

 Several test videos are included in the edge and remote servers. To test the RL-based VoD service, we use a standard video playback client (smartphone or laptop). The client asks for video content and, based on the cache status on the edge server, is served by the edge or remote server. 
 
 The private 5G network and VoD streaming service are deployed in Barcelona (CTTC servers). During the demo, we can monitor the delay, serving server, video playback quality, and after several tests, the effectiveness of ML-based caching in adapting to the new traffic pattern.

Finally, we made a short video clip, which explains the different parts of the system architecture (Fig. \ref{fig:implementation}), VoD streaming, and RL-based caching. The video file also includes a short demo of VoD service with two test contents and can be found in \cite{video_demo}.

%------------------------------------------------------------
%------------------------------------------------------------
\section{KPIs}
The relevant KPIs for the edge-cache are cache hit ratio, storage usage, and effective content (share of contents that diccount for $80\%$ of traffic). We define 4 scenarios based on content popularity and storage usage. The KPIs for each scenario are reported in Table. \ref{KPIs_table}. We can see that cache hit ratio of 0.8 can be achieved in all the scenarios by adding storage. One can even improve the cache hit ratio further, but the trade-off between backhaul traffic cost and storage cost in the edge should be considered.

\begin{table}
	\renewcommand{\arraystretch}{1}
	\caption{KPIs for different scenarios.}
	\label{KPIs_table}
	\centering
	\begin{tabular}{|l|c|c|c|l| } 
		\hline
		KPIs &  \thead{Storage
			\\(\% of total)} & \thead{Cache hit \\ratio} & \thead{effective\\contents} \\ 
		\hline\hline
		1  & $10\%$& $0.74$& $5\%$ \\  
		\hline
		2 &$20\%$& $0.82$& $5\%$ \\
		\hline
		3 &$20\%$& $0.75$&$10\%$ \\ 
		\hline
		4  &$30\%$& $0.81$&$10\%$ \\
		\hline
	\end{tabular}
\end{table}
%\begin{table}
%	\renewcommand{\arraystretch}{1}
%	\caption{KPIs for different scenarios.}
%	\label{KPIs_table}
%	\centering
%	\begin{tabular}{|l|c|c|c|c|l| } 
%		\hline
%		KPIs & \thead{Delay\\ (mean/std ms)} & \thead{Storage
%			\\(\% of total)} & \thead{Cache hit \\ratio} & \thead{effective\\contents} \\ 
%		\hline\hline
%		1 &$80,20$& $10\%$& $0.74$& $5\%$ \\  
%		\hline
%		2 & $80,20$&$20\%$& $0.82$& $5\%$ \\
%		\hline
%		3 & $80,20$ &$20\%$& $0.75$&$10\%$ \\ 
%		\hline
%		4 & $80,20$ &$30\%$& $0.81$&$10\%$ \\
%		\hline
%	\end{tabular}
%\end{table}
%We should also consider that the discussed approach is only valid when the edge-server and remote-server are part of the same business entity. Otherwise, further investigation needs to be done to accommodate encrypted traffic. This is an active research topic, and several solutions such as blind-cache \cite{eriksson2017blind}, cache proxy, and crypto-cache \cite{CryptoCache2017} are proposed.
%------------------------------------------------------------
%------------------------------------------------------------
\section{Summary}
We implemented a video-on-demand service using RL-based cashing for 5G networks. Firstly, the RL-based solution outperforms the traditional approach like LFU based caching. Secondly, the HLS-based video-on-demand service is implemented and managed by OSM, making it a good fit for the new design paradigm of the 5G networks. Finally, the reported KPIs demonstrate satisfactory performance cache hit ratio.
%------------------------------------------------------------
%------------------------------------------------------------
\section*{Acknowledgment}
This work is founded by the H2020 5GSolutions (Grant Agreement no.856691). Also, discussions with Luca Vettori and Miquel Payaró from Centre Tecnològic Telecomunicacions Catalunya (CTTC), are gratefully acknowledged

%-----------------------------------------------------------------------------------------------------

\bibliographystyle{IEEEtran}
\bibliography{ref}

\end{document}